\documentclass[aps,twocolumn,floatfix,showpacs,amsmath,superscriptaddress]{revtex4}
\usepackage{graphicx}
\newcommand{\be}{\begin{equation}}
\newcommand{\ee}{\end{equation}}
\newcommand{\bea}{\begin{eqnarray}}
\newcommand{\eea}{\end{eqnarray}}

\begin{document}


\title{Dynamics of Three Agent Games}


\author{Muhittin Mungan}\email[]{mmungan@boun.edu.tr}
\affiliation{Bo\~{g}azi\c{c}i University Department of Physics \\ Bebek 34342, Istanbul\\ 
Turkey}
\affiliation{The Feza G\"ursey Institute,\\
P.O.B. 6, \c Cengelk\"oy, 34680 Istanbul, Turkey}
\author{Tongu\c{c} Rador}\email[]{tonguc.rador@boun.edu.tr}
\affiliation{Bo\~{g}azi\c{c}i University Department of Physics \\ Bebek 34342, Istanbul\\ 
Turkey}


\date{\today}

\begin{abstract}

We study the dynamics and resulting score distribution of three-agent games where after each competition a single agent wins and scores a point. A single competition is described by a triplet of numbers $p$, $t$ and $q$ denoting the probabilities that the team with the highest, middle or lowest accumulated score wins. We study the full family of solutions
in the regime, where the number of agents and competitions is large, which can be regarded as a hydrodynamic limit. Depending on the parameter values $(p,q,t)$, we find 
six qualitatively different asymptotic score distributions and we also provide a qualitative understanding of these results. We checked our analytical results against 
numerical simulations of the microscopic model and find these to be in excellent agreement. 
The three agent game can be regarded as a social model where a player can be favored or disfavored for advancement, based on his/her accumulated score. It is also possible to decide the outcome of a three agent game through a mini tournament of two-agent competitions among the participating players and it turns out that the resulting possible score distributions are a subset of those obtained for the general three agent-games. We discuss how one can add a steady and democratic decline rate to the model and present a simple geometric construction that allows one to write down the corresponding score evolution equations for $n$-agent games. 
\end{abstract}


\maketitle

\section{Introduction}

The use of physical-mathematical modeling to describe phenomena involving biological, social, political, economical and many other systems is becoming more relevant and appreciated and has become an active area of research. One reason this is possible is because all of these systems generally admit a simple interpretation: a sum of agents and a set of rules to describe their interaction \cite{pap1}-\cite{pap3}. A reasonable approach to model the interactions is to involve a finite number of agents at a time and via the rules decide a winner which gains one unit of the attribute used to compare the agents.  The interaction may represent, a competitive game for wealth \cite{wel1}-\cite{wel4}, trophies in sports \cite{sp1},\cite{sp2} , opinion dynamics \cite{op1}-\cite{op3}, idea or rumor propagation \cite{rum1}-\cite{rum3}. One can also contemplate the emergence of social hierarchies from such models \cite{soc1}-\cite{soc5}. On the other hand, one can contemplate situations in which the 
game is not purely competitive and where the weaker players might have a higher chance of winning. Such models may admit solutions that can be useful for making analogies with welfare systems, for example.

In this paper we analyze an extension of a recently introduced two-agent model \cite{soc1}-\cite{soc5} to three-agent interactions \cite{3pap}, where a single competition involves three individuals at a time. The winner in a competition is determined by three numbers $(p,t,q)$ describing the winning probability of the agent with largest, middle and lowest points, respectively. This model has been studied in detail for the three special cases $(1,0,0)$, $(0,1,0)$ and $(0,0,1)$ in \cite{3pap}. The aim of the present paper is to study the model for the entire phase space of parameters, bounded only by the conservation of probability, $p+t+q=1$. We find that the full set of resulting solutions contains qualitative differences that go  beyond the three special cases considered in \cite{3pap}. Three-agent games can be motivated by the observation that in social interactions generally more than two agents are involved, such as, in a job application or companies competing for a contract.

The organization of the paper is as follows. In section II we present the model and its mathematical interpretation, yielding the equations governing the point distribution of the agents. In section II we also present a simple geometric method to generalize the model to $n$-agent competitions. Section III is reserved to  the study of the solutions and the phase space, where the solutions are also interpreted in terms of social structures in a society of agents characterized by their scores. We also discuss in that section the resolution of a subset of three-agent games in terms of a mini-tournament involving two-agent interactions. We finish the section by introducing a decline rate for the agents to describe the loss of fitness due to inactivity as advocated in \cite{soc5}. The last section is reserved for a short discussion and comments on possible interesting extensions of the model.

\section{The Model}

In the model we would like to study a competition is described as follows: We first pick three agents out of a collection of $N$. We advance only one agent, based on their accumulated points prior to the competition. Let the first, second, and third agents have scores $x$, $y$ and $z$, respectively. Then, the  agent with the largest score will increase its score by one unit with probability $p$, the one with the smallest score with probability $q$ and the one in the middle with probability $t$. The situation with equal scores are evaluated on the basis of equal likelihood. The initial condition is that at the beginning all agents start with a zero score.  

Assuming an ordering of points is made so that one has $x\geq y\geq z$, we can cast the rules of the competition as

\begin{subequations}{\label{eq:micro}}
\begin{eqnarray}
\left(x>y>z\right)    &\Longrightarrow&      \left(p, \;t, \;q\right)\\
\left(x=y>z\right)    &\Longrightarrow&      \left(\frac{p+t}{2},\; \frac{p+t}{2},\; q\right)\\
\left(x>y=z\right)    &\Longrightarrow&      \left(p,\; \frac{t+q}{2},\; \frac{t+q}{2}\right)\\
\left(x=y=z\right)    &\Longrightarrow&      \left(\frac{1}{3},\; \frac{1}{3},\; \frac{1}{3}\right)
\end{eqnarray}
\end{subequations}

\noindent Here the lists on the right represent the winning probabilities of the teams with points listed on the left. That after every game one agent surely advances requires the normalization of the probabilities

\be
p+t+q=1\;.
\ee

From this microscopic model we can read out the change of the number of teams in a particular score range. Let us denote by $f_{x}$ the fraction of teams having score $x$. After a competition some teams might leave this region and some teams might enter it by winning a competition in each case. This suggests the following local conservation law,

\be{\label{eq:1}}
\frac{df_{x}}{d\tau}=\sum_{y,z} \left( f_{x-1}\;W_{x-1,y,z}-f_{x}\;W_{x,y,z}\right) \;f_{y}\;f_{z}\;.
\ee

\noindent Here $W_{x,y,z}$ denotes the probability that the team with $x$ points will win and  the microscopic rules in Eq.~(\ref{eq:micro}) completely define what it is. The right hand side is cubic in $f$, since we are picking three agents out of the collection, and the probability to pick a team with a given point $x$ is $f_{x}$. 

Since 

\be
\sum_{x}f_{x}=1, 
\ee

\noindent it is immediate that equation Eq.~(\ref{eq:1}) also implies, as it should, the global conservation of the total number of teams, as can be checked by performing a sum over 
$x$. 

The time variable $\tau$ in Eq.~(\ref{eq:1}) has an arbitrary scaling which can be compensated by an overall factor in the definition of $W$. The natural scale is such that the average points of teams is given by,

\be
\bar{x}(\tau)\equiv\sum_{x=0}^{\infty}x\;f_{x}=\frac{\tau}{3}\;,
\label{eqn:xave1}
\ee

\noindent meaning that (on average) each team participates in a single game during each round $\tau$ of games. As only one of the participating teams in a game wins and advances its score by one,  Eq.~(\ref{eqn:xave1}) follows. 

The presence of sums over the discrete indices on the right hand side of Eq.~(\ref{eq:1}), 
results in a coupled set of differential equations. These can be further simplified by defining

\be
F_{x}\equiv\sum_{x'=0}^{x-1}f_{x}
\ee

\noindent so that 

\be
f_{x}=F_{x+1}-F_{x}\;.
\ee

\noindent Summing (\ref{eq:1}) over $x$ we obtain

\be\label{eq:2}
\frac{dF_{x}}{d\tau}=-f_{x-1}\sum_{y,z}W_{x-1,y,z}\;f_{y}\;f_{z}\;,
\ee

\noindent where the surface term at $x=-1$ has been ignored due to the fact that no team can have negative points. 

Now note that 
\be\label{eq:sum}
f_{x-1}\sum_{y,x}W_{x-1,y,z}\;f_{y}\;f_{z}
\ee 

\noindent yields the probability that a team with score $x-1$ will win any possible choice of single competition with two other teams. Since $W$ depends on the ordering of points, one can easily work out the sum from the rules, Eq.~(\ref{eq:micro}), and Eq.~(\ref{eq:1}) yielding

\begin{figure}[t]
\includegraphics[scale=0.75]{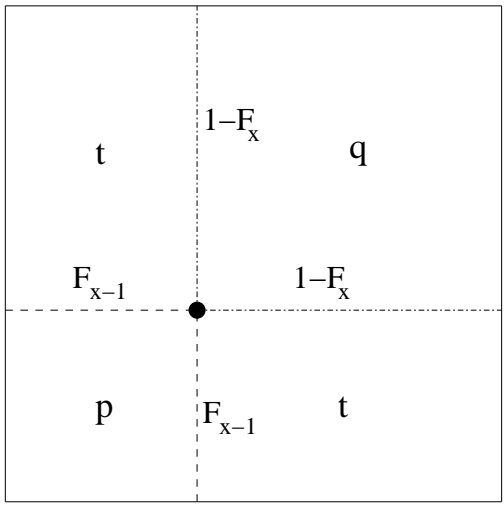}
\caption{{\label{fig:eq1}} The geometric representation of the sum in Eq.~(\ref{eq:sum}). The filled circle represents the team with point $x-1$. The dashed and dot-dashed lines represent the number of teams with points strictly less or greater than $x-1$. The sum over the areas will bring the bulk interaction whereas the sum over the interfaces represent the cases where two teams have equal points. The dot gives the term when all the teams have equal points. This construct is easily generalized to n-agent games as discussed in the text.}
\end{figure}
 

\bea\label{eq:harbi}
\frac{dF_{x}}{d\tau}=&&-f_{x-1}\left[ pF_{x-1}^{2}+q(1-F_{x})^{2}+2tF_{x-1}(1-F_{x})  \right] \nonumber \\
&&-2\frac{(p+t)}{2}f_{x-1}^{2}F_{x-1}\nonumber \\
&&-2\frac{(t+q)}{2}f_{x-1}^{2}(1-F_{x})\nonumber\\
&&-\frac{1}{3}f_{x-1}^3
\eea

\noindent The terms on the first line of Eq.~(\ref{eq:harbi}) represent the bulk of interactions between three players with different scores. The second and third lines represent the cases where two players have identical scores $x$ and the last term represents the case when all the teams have equal score.

Alternatively, the terms on the right hand side of Eq.~(\ref{eq:harbi}) can be obtained 
by invoking a geometric construction: Imagine all the players aranged on a line ordered by their scores and let the $x$ and $y$ axis formed in this way represent the possible opponents of a player with a given score and let two perpendicular axes formed in this way represent the possible opponents of a player in the competition. The situation is represented graphically in Fig.~\ref{fig:eq1}, where the filled circle represents a player with score $x-1$. The dashed and dot-dashed lines represent the number of opponents  with scores strictly less or greater than $x-1$. The sum over the areas constitutes the bulk of the match-ups in which all three players have different scores, whereas the sum over the interfaces represent the cases where two teams have equal scores. The dot gives the term when all the teams have equal scores.

In a microscopic simulation of the system the interface terms are the most important only at early times, since at the beginning all teams have zero points. In fact, the most important contribution at the very beginning is described by the $f_{x-1}^{3}$ term in Eq.~(\ref{eq:harbi}). The effect of this term is to create random imbalances in the point distribution which in time will make the $f_{x-1}^{2}$ terms more relevant. These imbalances are further emphasized by the dynamics of the bulk terms as prescribed by the set of winning probabilities $\left(p,t,q\right)$. Therefore as time goes by, in a thermodynamic limit where the number of teams ranges to infinity, the majority of the contributions to the dynamics will be governed by the bulk terms. On the other hand as time goes by, almost every team will accumulate a certain number of points which, in general, will be larger than a single point. These considerations allow one to go to a continuum limit where the differences are expanded in terms of the derivatives. A first order approximation, where one considers only the bulk terms, results in the following,

\be{\label{eq:cont1}}
\frac{\partial F}{\partial \tau}=-\frac{\partial F}{\partial x}\; G'(F)\;,
\ee

\noindent with 

\be{\label{eq:cont2}}
G'(F)\equiv pF^{2}+2tF(1-F)+q(1-F)^2\;.
\ee

\noindent As thoroughly studied in \cite{soc5}, the contribution  of the higher order derivative terms, that is the interface terms, become negligible in the infinite time limit.
We will refer to this regime as the hydrodynamical limit. 

The fact that in the hydrodynamical limit the interface terms become negligible is also corraborated by the fact that they do not contribute to the asymptotic time dependence of the average score. In fact, we have

\be
\frac{d\bar{x}(\tau)}{d\tau}=\frac{d}{d\tau}\int_{0}^{\infty}dx\;x\;f=-\int_{0}^{\infty}dx\frac{\partial F}{\partial \tau}\;,
\ee

\noindent and using Eq.~(\ref{eq:cont1}) we find

\be
\frac{d\bar{x}(\tau)}{d\tau}=\int_{0}^{\infty}dx\;\frac{\partial F}{\partial x}G'(F)\;.
\ee

\noindent Noting that $F(x=\infty)=1$ and that in the large time limit $F(x=0)=0$ (since a team with zero points will happen with vanishing probability in the limit of large times) we get

\be
\frac{d\bar{x}(\tau)}{d\tau}=\int_{0}^{1}dF\;G'(F)=\frac{p+t+q}{3}=\frac{1}{3}\;.
\ee

\noindent which implies that asymptotically $\bar{x}(\tau)=\tau/3$.

\subsection{Generalization to n+1-agent games}

The geometric construction leading to Eq.~(\ref{eq:1}) can easily be generalized to n+1-agent games. For simplicity, we will consider the form of the equation only in the hydrodynamical limit, where one considers solely the bulk effects. One can easily generate the interface interactions, if needed. The construction is simply picking a point in an $n$-cube and splitting its volume by $n-1$ planes passing through this point that are orthogonal to each other and also to the sides of the cube. As is well known, the number of resulting equivalent $n$-volumes is given by the binomial coefficients. The dynamics of a $n+1$-agent game, in the hydrodynamical limit, is therefore given by

\be
\frac{\partial F}{\partial \tau}=-\frac{\partial F}{\partial x}\; G'(F)\;,
\ee

\noindent with

\be
G'(F)=\sum_{k=0}^{n}\left(
\begin{array}{l}
n \\
k 
\end{array}
\right)p_{k}(1-F)^{k} F^{n-k}\;.
\ee

The probabilities $p_{k}$ for the player with the $k$th highest score to win obey of course

\be
\sum_{k=0}^{n}p_{k}=1\;,
\ee

and the mean score thus evolves as

\be
\bar{x}(\tau)=\frac{\tau}{n+1}\;.
\ee

\section{Analysis of the Model}

In this section we first introduce the quantitative methods to analyze our model. This is followed by a study of the full parameter range bound by $p+t+q=1$ resulting in a phase diagram of the possible types of solutions for the model. For every regime in the phase diagram we will present a numerical simulation of the microscopic model and give a qualitative understanding of the hydrodynamic limit in that case. The score distributions are next  analysed in terms of their implied social structure through ranking. A three agent-game can be obtained by a mini-tournament among the players based on a two-agent approach. We show that these cases constitutes a curve on the phase plane, {\it i.e.} they form a subset of solutions to the genuine three agent game. We finish this section by discussing the effect of adding a decline rate for the agents.

\subsection{The scaling Ansatz}

The form of the differential equation for $F$, Eq.~(\ref{eq:cont1}), suggests a scaling solution ansatz of the form

\be
F(x,\tau)\to F(z\equiv\frac{x}{\tau})\;.
\ee

\noindent which yields 

\be
\frac{{\rm d}F}{{\rm d}z}\left[-z+G'(F)\right]=0\;.
\ee

\noindent The scaling equation thus obtained means that one either has $F={\rm constant}$ or $G'(F)=z$. The boundary conditions are $F(z=0)=0$ (since in the infinite time limit the probability to have a team with zero points vanishes) and $F(z=1)=1$ (because the maximum possible points a team can accumulate in time $\tau$ is simply $\tau$ in our normalization of the time variable). Furthermore, since the definition of $F$ means that it has to be an increasing function of $x$ and hence $z$ we have either ${\rm d}F/{\rm d}z=0$ or ${\rm d}F/{\rm d}z>0$.

For a three-agent game $G'(F)$ is quadratic. Thus one can generally expect regions of parameters where $d^{2}{F}/dz^{2}$ is positive or negative which implies one can also expect a critical point of a crossover where $d^{2}F/dz^{2}$ vanishes. 

On the other hand, the possibility  of patching solutions of the two mentioned forms already suggests that the derivative of $F$ need not be continuous. In fact, $F$ itself can have  discontinuities, resulting in  Dirac-delta singularities in its derivative, the score distribution function and implying that a finite fraction of teams evolves with the same winning rate. At this point we would like to reemphasize that the contribution of the interface and hence the higher order derivative terms will tend to smooth out these discontinuities but that their overall effect is negligible in the inifinite time limit. We refer the interested reader to \cite{soc5}, where the minute effect of these terms were studied in detail for the two-agent version of the model presented.

The main problem, therefore, is to unambiguously obtain the solution for any given  $z$ in the hydrodynamical limit. This is most readily carried out by noting that Eq.~(\ref{eq:cont1}) is in the form of a hyperbolic conservation law and applying the theory of weak solutions of hyperbolic conservation laws, also know as the {\it method of characteristics}. In the next subsection we therefore turn to a brief description of the method of characteristics and then proceed to obtain the solution of Eqs.~(\ref{eq:cont1}) and (\ref{eq:cont2}) for the full parameter range $p + q+ t = 1$. For a more detailed account of the theory of hyperbolic 
PDEs and conservation laws, we  refer the reader to \cite{Leveque} and references therein.  

\subsection{The Method of Characteristics}

We are looking for a solution of  

\be{\label{eqn:Fgen}}
\frac{\partial F}{\partial \tau}=-\frac{\partial F}{\partial x}\; G'(F)\;,
\ee

\noindent subject to the initial condition 

\begin{equation}
F(x,0) = \left \{ \begin{array}{ll}  0 & x<0, \\ 1 & x \ge 0. \end{array} \right.
\label{eqn:initial_data}
\end{equation}

Given (\ref{eqn:Fgen}), we define its characteristics as the curves $x(\tau)$ in the 
$x-\tau$ plane on which $F(x,\tau)$ remains constant. It can be shown that these 
curves are given by the {\em characteristic equation}

\begin{equation}
x(t) = x_0 + \tau G'(F(x_0,0))
\label{eqn:characteristics}
\end{equation}

\noindent and thus $G'(F(x_0,0))$ is the speed of the characteristic emerging from the point $x_0$.

An implicit solution can therefore be found as $F(x,\tau) = F(x_0,0)$, where  for a given $(x,\tau)$, $x_0$ is determined from the characteristics, as defined in (\ref{eqn:characteristics}). 

Depending on the initial conditions and the form of $G'(F)$, the characteristics can  intersect, giving rise to multiple-valued points that are resolved by discontinuities (shocks), spread out giving rise to continuous solutions (rarefaction waves), or  do both. 
Even with smooth initial data the solution can develop discontinuities in a finite time. 

One therefore seeks weak solutions, which for our purposes can be defined as solutions to the "viscous" equation

\begin{equation}
\frac{\partial F}{\partial \tau} + \frac{\partial F}{\partial x} G'(F) = \epsilon \frac{\partial^{2}F}{\partial x^{2}}\;,
\label{eqn:Fgen_visc}
\end{equation}

\noindent in the limit that $\epsilon \rightarrow 0^{+}$. It is well-known that this prescription is equivalent to seeking solutions such that 

\begin{equation}
\int_{0}^{\infty}  \int_{0}^{\infty} \chi(x,\tau) \left [  \frac{\partial F}{\partial \tau} + G'(F)  \frac{\partial F}{\partial x} \right ] d\tau \;dx  = 0
\label{eqn:weak}
\end{equation}

\noindent for any continuously differentiable function $\chi(x,\tau)$ with compact support \cite{Leveque}. This condition allows one to obtain weak solutions without having to solve  the "viscous" equation.

\subsubsection{Application to the two-agent game}

In order to illustrate the method of characteristics we consider first the two-player game. In this case we have

\be
G'(F) = q + (p-q) F\;,
\ee

\noindent with $p+q = 1$ and the same initial data, Eq.~(\ref{eqn:initial_data}). The speed of the characteristics are thus $G'(0) = q$ and $G'(1) = p$.

For $p>q$ the characteristics spread out and we have $F(x,\tau) = 1$ for $x>p\tau$ and 
$F(x,\tau) = 0$ for $x<q\tau$, while in the region $q\tau \le x \le p\tau$ the weak solution results 
in

\begin{equation}
G'(F) = \frac{x}{\tau} \equiv z \; .
\end{equation}

In terms of the scaling variable $z$ we therefore find 

\be
F(z) = \frac{z - q}{p-q}
\ee

\noindent for $q\leq z\leq p$.

The interpretation is that the strongest teams are becoming stronger, increasing their scores at a rate $p$, while the weakest teams can only increase their scores at a lower rate  $q$. The score rate of the majority of teams lies in between these extreme cases and turns out to be uniformly distributed which is mandated by the fact that in this case $G'(F)$ is linear.

For $p<q$, the characteristics intersect. This means that once a team starts winning a series of games its increased score makes it less likely to win against most of the weaker teams, causing its score rate to decline. In the limit of very large scores, the average score rate of a team is $1/2$ (since only half of the teams participating in a round of matches win) and fluctuations around this average become increasingly less (of the order of $\tau^{-1/2}$), leading to the shock solution in the infinite time, {\it i.e. hydrodynamical}, limit.

For any G(F), the shock speed is given by the Rankine-Hugoniot jump condition \cite{Leveque},

\begin{equation}
 v = \frac{G(F_l) - G(F_r)}{F_l - F_r } = \frac{\Delta G}{\Delta F}, 
\end{equation}

\noindent where $F_l$ and $F_r$ are the values of $F$ immediately to the left and right of the 
discontinuity. 

For the two-player game, when $q>p$, $F_l = 0$ and $F_r = 1$, while $G(F) = q F + (p-q)F^2\;/2$ and the result $v=1/2$ indeed follows. 

\subsection{The Three-Player Game}

Depending on $p,t,$ and $q$, the function $G(F)$ is not necessarily convex in the 
interval $F \in [0,1]$ meaning that $G'(F)$ is not a monotonously increasing or decreasing function as was the case in the two-player game. As we have mentioned before whenever $G'(F)=z$ is part of the solution one can expect $F$ to be concave up or down in $z$. When  $G'(F)$ is non-monotonous this implies that $F$ has discontinuities, {\it i.e.} shocks. For  general, $p,t,$ and $q$, the resulting solution of $F$ turns out to contain a continuous segment given by $G'(F)=z$ as well as a single discontinuity. We find the following regimes

\begin{itemize}
\item {\bf $C^{-}$:}  $p>t \ge q$ and $\tau < 1/3$, 
\item {\bf $C^{0}$:}  $t = 1/3 > q$, 
\item {\bf $C^{+}$:}  $t > 1/3$ and $p \ge t$,
\item {\bf $C^{+}_{S}$:}  $p< t$ and $q \le 1/3$, 
\item {\bf $S$:}  $q > 1/3 > p$,
\item {\bf $C^{-}_{S}$:}  $q > t $ and $p > 1/3 $.
\end{itemize}

\begin{figure}[t]
\includegraphics[scale=0.333]{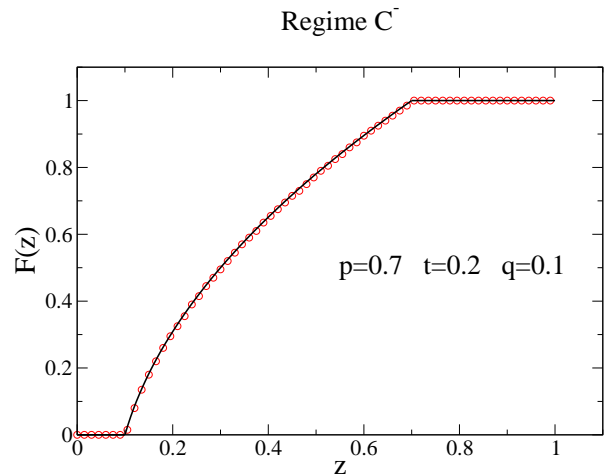}\\
\caption{A representative graph for the solution in regime $C^{-}$. The circles represent the data from the simulation of the microscopic model. The solid line is the analytical solution.}
\label{fig:cminus}
\end{figure}

\begin{figure}[t!]
\includegraphics[scale=0.333]{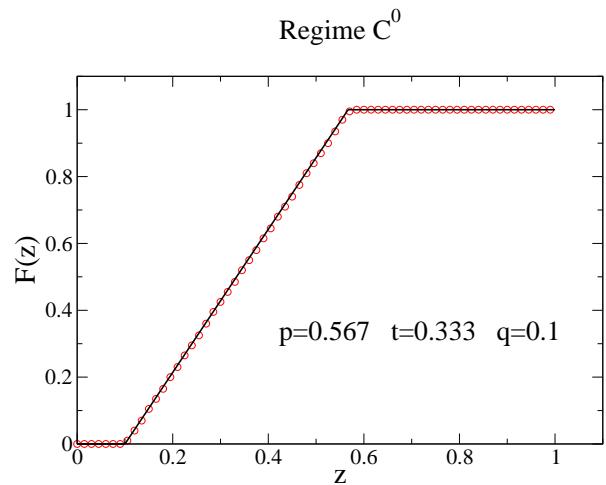}\\
\caption{A representative graph for the solution in regime $C^{0}$. The circles represent the data from the simulation of the microscopic model. The solid line is the analytical solution.}
\label{fig:czero}
\end{figure}

\begin{figure}[t!]
\includegraphics[scale=0.333]{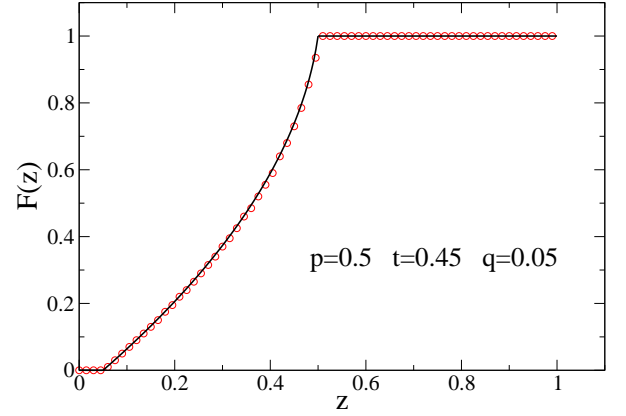}\\
\caption{A representative graph for the solution in regime $C^{+}$. The circles represent the data from the simulation of the microscopic model. The solid line is the analytical solution.}
\label{fig:cplus}
\end{figure}

\begin{figure}[t!]
\includegraphics[scale=0.333]{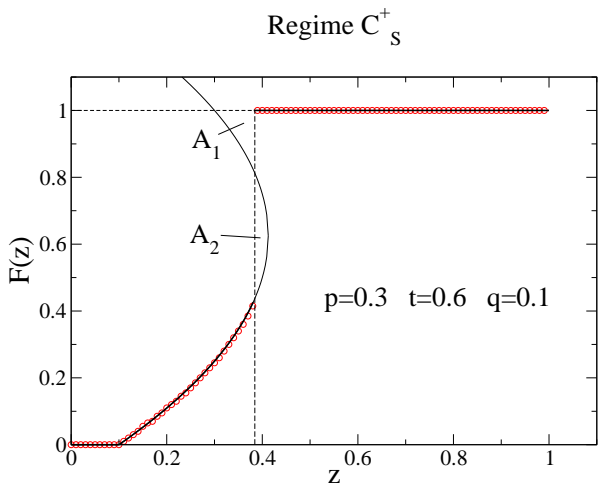}\\
\caption{A representative graph for the solution in the regime $C^{+}_{S}$. The circles represent the data from the simulation of the microscopic model. The solid line is the analytical solution. The Maxwell construction which determines the location of the shock requires that the areas $A_{1}$ and $A_{2}$ be equal.}
\label{fig:cSplus}
\end{figure}

\begin{figure}[t!]
\includegraphics[scale=0.333]{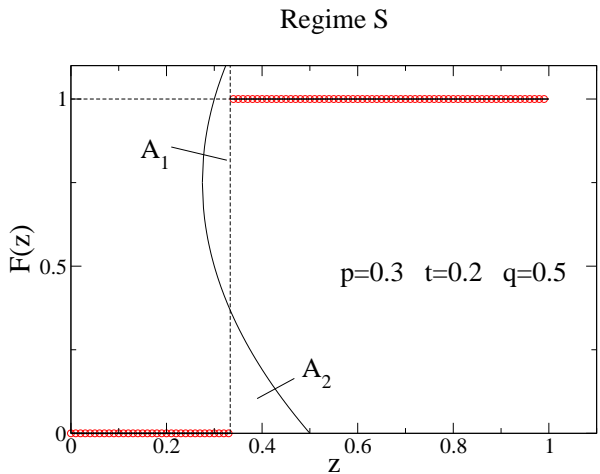}\\
\caption{A representative graph for the solution in the regime $S$. The circles represent the data from the simulation of the microscopic model. The solid line is the analytical solution. The Maxwell reconstruction to determine the location of the shock requires that the areas $A_{1}$ and $A_{2}$ are equal.}
\label{fig:S}
\end{figure}

\begin{figure}[t!]
\includegraphics[scale=0.333]{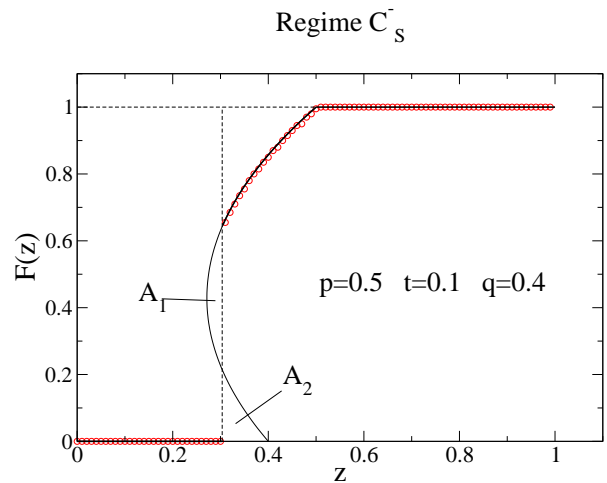}
\caption{A representative graph for the solution in the regime $C^{-}_{S}$. The circles represent the data from the simulation of the microscopic model. The solid line is the analytical solution. The Maxwell reconstruction to determine the location of the shock requires that the areas $A_{1}$ and $A_{2}$ are equal.}
\label{fig:cSminus}
\end{figure}

\noindent In this suggestive notation, which will be justified below, regimes $C^{+}$ and $C^{-}$ represent a solution $F$ which is concave up/down in $z$ respectively. The regime $C^{0}$ is the interface between the concave up and down solutions and has a vanishing second derivative. The label $S$ means that there is a shock form in the solution. So the pure $S$ regime means that the solution has the form of a single shock. That is,  a step function solution for $F$. The corresponding profiles are shown in Figs.~(\ref{fig:cminus})-(\ref{fig:cSminus}).

The solution in all six regions can thus be written in the general form 

\begin{equation}
F(x,\tau) = \left \{ \begin{array}{ll} 0 & z < z_l, \\
				\Phi(z) & z_l \le z \le z_r, \\  
			1 & z > z_r, 
\end{array} \right.
\label{eqn:genform}
\end{equation}
where, again, $z = x/\tau$.  
\noindent We also define the roots $\Phi_{\pm}(z)$ of the equation

\begin{equation}
G'(\Phi) = z
\end{equation}

\noindent as 

\begin{equation}
\Phi_{\pm}(z) = \frac{q - \tau \pm \left [ ( q - \tau)^2 + (1 -3\tau)(z-q) \right ]^{1/2}}
{1 -3\tau}
\end{equation}

\noindent Note in particular that 

\begin{equation}
\Phi_{\pm}(q) = \frac{q - \tau \pm \vert q - \tau \vert }{1 -3\tau}
\end{equation}

\noindent and 

\begin{equation}
\Phi_{\pm}(p) = \frac{q - \tau \pm \vert p - \tau \vert }{1 -3\tau}
\end{equation}


In regimes $C^{-}$, $C^{0}$ and $C^{+}$ the function $G'(F)$ increases monotonously from $q$ to $p$ as $F$ goes from 0 to 1. 
This results in a  continuous solution for $q \le z  \le p$. In regions with the label $S$ on the other hand, 
$G'(F)$ does not have a monotonous behavior and from the characteristic construction we 
see regions that are multiple-valued. These cases are resolved by the Maxwell (or equal-area) construction, as we will describe below, which is a consequence of the weak solution prescription, Eq.~(\ref{eqn:weak})\cite{Leveque}.

\vspace*{0.3cm}
\noindent {\bf Region $C^{-}$:  $p>1/3> \tau \ge q$}
\vspace*{0.3cm}

Here we have $\Phi_+(q) = 0$, $\Phi_+(p) = 1 $ and the solution $F(x,\tau)$ is given by Eq.~(\ref{eqn:genform}) with $\Phi(z) = \Phi_+(z)$, $z_l = q$ and $z_r = p$.

To understand the structure emerging from this solution, let us recall that the population density of agents $f$ is given by the first derivative of $F$. The fact that $F$ is a concave down and monotonously increasing function for this case implies that most of the agents are in the lower point range. Thus this situation makes a contrast to the two-agent game in the sense that there this distribution is uniform. One can argue that this is related to the fact that in this case of the three agent-game the ordering of winning probabilities constitutes a pure rich-get-richer competition and therefore once a team wins a game it actually beats {\rm two lower rank} teams. This is in agreement with the fact the the middle team wins with a probability less than $1/3$ and gives rise to the asymmetric first derivative at $F=0$ and $F=1$. The role of the value of $t$ relative to $1/3$ will become clearer when we study the region $C^{0}$ below.

At this point we would like to introduce two useful quantities in assessing the score  distributions emerging from the three agent game. The ratio of the number of the poorest agents to the richest agents is related to the score distribution and is given by

\be
\omega\equiv\frac{f(z_{l})}{f(z_{r})}=\frac{F'(z_{l})}{F'(z_{r})}\;\;.
\ee

Another useful quantity is the length of the score region over which the agents are distributed and which is associated with the total score shared by them, 

\be
\sigma\equiv\frac{z_{r}-z_{l}}{1}\;\;.
\ee

Since these metrics characterize the distributions of the score of agents within their society of players, we will refer to these metrics as "social" indices. In the solution regime $C^{-}$ we are considering here they are given by 

\begin{subequations}
\bea
\omega_{C^{-}}&=&\sqrt{1+\frac{(1-3t)(p-q)}{(q-t)^2}}>1\;,\\
\sigma_{C^{-}}&=&p-q\;.
\eea
\end{subequations}

\noindent That $\omega_{C^{-}}>1$ shows the bias in the distribution as opposed to the two agent game. This bias is most extreme in the   special case  where $q = t \Rightarrow p = 1 - 2t$,  where $\Phi$ becomes 

\begin{equation}
\Phi(z) = \left [  \frac{z-t}{1 -3t} \right ]^{1/2}
\end{equation}

\noindent and the score distribution $\omega$ diverges as a simple pole. This means that a great majority of the agents are near the poorest ones.

\vspace*{0.3cm}
\noindent {\bf Region $C^{0}$: $p>t = 1/3 > q$}
\vspace*{0.3cm}

We again have $z_l = q$, $z_r = p $ and $\Phi(z)=\Phi_{+}(z)$, which simplifies to

\begin{equation}
\Phi(z) = \frac{z-q}{p-q}
\end{equation}

\noindent for $q\leq z\leq p$. 

Note that the resulting profile 
is qualitatively the same as the one for the two player game with $q < p$. 
This is so, because the middle team in a three player set has a 1/3 chance of winning the game and thus the team's score does not induce any bias towards either to the rich or poor side of the spectrum of agents. So $1/3$ of the games can be considered to be completely redundant given  $p>t = 1/3 > q$. Thus being in the middle of a triplet of agents is just like a pure random walk. In this situation the social indices we have introduced before become

\begin{subequations}
\bea
\omega_{C^{0}}&=&1\;,\\
\sigma_{C^{0}}&=&p-q\;.
\eea
\end{subequations}

\noindent that is, the number of the richest agents is the same as the number of poorest agents. Note that in this case the number of agents with any given score is the same. This is related to the fact that in a three-agent game $G'(F)$ is at most quadratic  and that $F$ has to be increasing.

\vspace*{0.3cm}
\noindent {\bf Region $C^{+}$: $p\ge t > 1/3>q$}
\vspace*{0.3cm}

We still have $z_l = q$, $z_r = p$ and $\Phi(z) = \Phi_{+}(z)$. The social indices become

\bea
\omega_{C^{-}}&=&\sqrt{1-\frac{\mid1-3t\mid(p-q)}{(q-t)^2}}<1\;,\nonumber\\
\sigma_{C^{-}}&=&p-q\;.
\eea

On the line $t = p \Rightarrow q = 1 - 2t$, we have in particular
\begin{equation}
\Phi(z) = 1 - \left [1 + \frac{z-q}{1- 3q }\right ]^{1/2}\;.
\end{equation}

\noindent In this extreme case we have $\omega_{C^{-}}=0$ meaning that the majority of the agents are condensed near the rich side $z=p$. Note that the vanishing of $\omega_{C^{-}}$ is as that of a power-law.

We therefore see the common property of the regions without a shock front: In all three regimes one has the ordering of probabilities $p\ge t\ge q$ yielding a generally competitive game. The relative position of $t$ with respect to $1/3$ determines whether one has a mild accumulation of the agents towards the rich or poor side. When $t=1/3$ this ratio is equal. We present representative graphs of the regimes $C^{-}$, $C^{0}$ and $C^{-}$ in Figures~\ref{fig:cminus}, \ref{fig:czero} and \ref{fig:cplus}, respectively. The figures compare for a certain choice of parameters, both the score distribution obtained from  simulation of the microscopical competitions as well as our analytical results in the hydrodynamical limit.

\vspace*{0.3cm}
\noindent{\bf Region $C^{+}_{S}$: $t>p$ and $q<1/3$}
\vspace*{0.3cm}

In this parameter regime $G'(F)$ is not a monotonously increasing function in the interval $[0,1]$. This leads to a formation of a shock front at $z_r$, where the latter is 
determined by the Maxwell equal area construction: The general solution in this 
region is again of the form  (\ref{eqn:genform}) with $z_l = q$. but $z_{r}$ the location of the shock is determined by the equal area rule

\begin{equation}
G'[\Phi(z_r)] = \frac{G[1] - G[\Phi(z_r)]}{1 - \Phi(z_r)}
\end{equation}

\noindent After a little bit of algebra one finds that 

\begin{equation}
\Phi(z_r) = \frac{1}{2}\;\frac{1 - 3q}{3t - 1}
\end{equation}

\noindent and using $G'[\Phi(z_r)] = z_r$ we obtain

\begin{equation}
z_r = q + \frac{1}{4} \; \frac{3q -1 }{1 - 3t} \; (4t - q - 1)
\end{equation}

Thus we see that at $z_r$ we have a discontinuity that jumps from 
$F = \Phi(z_r)$ to $F = 1$. The resulting $\Phi$ profile is 
$\Phi(z) = \Phi_{+}(z)$, for $z_l \le z \le z_r $ followed by a shock discontinuity at
$z_r$.

For a single discontinuity, as above, it is clear that due to the equal area construction resulting in the vertical segment, the area under the graph $G'(F)$ remains 
unchanged and one has

\begin{eqnarray}
\frac{1}{3} &=& G(1) - G(0)  = \int_{0}^{1} G'(F) \; {\rm d} F  \nonumber \\
&=& \left [ 1-\Phi(z_r) \right ] G'[\Phi(z_r)]
 + \int_{0}^{\Phi(z_r)} G'(F) \; {\rm d} F,  
\end{eqnarray}

\noindent independent of $p,q$ and $t$, since it is the rate at which the total number of points 
accumulate as each team participates in a three player game. In fact, the equal area 
construction resolves a region with multiple valued points by a discontinuity such that 
the total conserved quantity remains unchanged. Thus for 
a single discontinuity this construction is equivalent to conserving the quantity over the whole domain as utilized in \cite{3pap}. 

On the other hand, when one has four or higher agent games, 
depending on the winning probabilities, the resulting profiles can have multiple shocks 
separated by rarefaction waves. In that case a global conservation constraint would not 
be sufficient to determine all shock locations, and one would have to resort to applying the 
equal area construction to each region where the profile is multiple valued. 

One can qualitatively understand the reason for a shock front to the right of the wealth span as follows. First of all, due to the fact that $t>p$, having the highest score in a competition is a disadvantage since a team in the middle of a triplet is favored. This results in a sort of deceleration mechanism for the propagation rate of  teams with high scores and consequently a fraction of the total agents condense at the high score side of the spectrum: they are the richest agents in the society. It is clear that such a case is an attractor solution since when it is formed it is not destroyed. That is, if an agent in this shock wins a game it will be disfavored in the future games resulting in a loss of its point. Conversely, if an agent in the shock looses a game it will be favored in a future game over a team in the shock resulting in its return to the shock region. 

The region with a continuous $F$ to the left of the shock can be understood easily if we remember that $q<1/3$ implies $q<(t+p)/2$, regardless of how $p$ and $q$ are ordered. Let us consider a game with two teams from the shock region and one from below, such a game constitutes a great majority of possible types of games if there is a shock region. In this case the probability of a win for the teams in the shock is given by $(t+p)/2$ and for the other by $q$. This means that the lower point team is disfavored altogether and the criterion for this is just $q<1/3$, resulting in a continuous population density for regions below the shock.

Due to the discontinuity in $F$ one has to exercise a little more care in implementing the social index $\omega$ we have introduced before. Since in this case
we have

\[
F(z)=\Phi_{+}(z)+\left[1-\Phi_{+}(z_{r})\right]\Theta(z-z_{r})
\]

there will be a strong singularity in $\omega$

\begin{subequations}
\bea
\omega_{C^{+}_{S}}&=&\frac{{\rm const}}{\delta(0)}\;,\\
\sigma_{C^{+}_{S}}&=&\frac{4t-q-1}{4}\left(\frac{1-3q}{3t-1}\right)\;.
\eea 
\end{subequations}

\noindent So $\omega$ vanishes much more strongly (and for all parameters) then in the particular case presented for the regime $C^{+}$ where the vanishing was determined by a simple zero. One, if willing, can define a better index which does not use the derivative of $F$. For instance, considering the agents in the left/right third of the distributed wealth $z_{r}-z_{l}$ as poor (slower moving agents) and rich (faster moving agents), respectively, a smeared index can be defined as

\be
\tilde{\omega}\equiv\frac{F(z_{l}+\frac{\sigma}{3})}{1-F(z_{r}-\frac{\sigma}{3})}\;,
\ee

\noindent which we will not pursue here.

\noindent A representative graph including an exact numerical simulation of the microscopic model and the analytical construction is presented in Fig.~\ref{fig:cSplus}.

\vspace*{0.3cm}
\noindent {\bf Region $S$: $q \ge 1/3 > p$}
\vspace*{0.3cm}

In this region $G'[F]$ is monotonously decreasing with $F$. The profile $\Phi$ is 
obtained from the equal area construction, resulting in a discontinuity at $z^{*}$ that covers 
the whole interval $[0,1]$ of $F$, {\it i.e.} a unit step. The shock speed is found to be 1/3 so that 
$z_l = z_r = z^{*}=1/3$. Note that when $p > t $ and $p \le 1/3 $, $G'[F]$ is not monotonously decreasing anymore in $F$. However the 
equal area construction still leads to a discontinuity at $z$ covering the whole interval of $F$ so that the shock speed is again 1/3 and thus $z^{*} = 1/3$. 

The social indices become $\omega_{S}=1$ and $\sigma_{S}=0$, reemphasizing the fact that in this case all agents share the same wealth. 

The meaning of a single shock becomes clear if we note that $q>p$ impliess that the lower score teams will eventually catch up with the teams with highest scores. The other condition $q > 1/3$ implies $q>(t+p)/2$ which means that the lower rank teams will also be able to catch up with a shock to the right of the wealth span. Note that $q>p$ does not necessarily imply $q>t$, although there is a subregion consistent with this condition. The condition $q>(t+p)/2$ ensures that even if $t>p$ and $t>q$, whenever there is a shock to the right as argued in our discussion of the region $C^{+}_{S}$, the lower rank teams will eventually reach that shock resulting in a single discontinuous front for the solution.

\noindent A representative graph including an exact numerical simulation of the microscopic model and the analytical construction is presented in Fig.~\ref{fig:S}.

\vspace*{0.3cm}
\noindent {\bf Region $C^{-}_{S}$: $q > t $ and $p > 1/3 $}
\vspace*{0.3cm}

In this parameter regime $G'(F)$ is not a monotonously increasing function in the interval $[0,1]$. This leads to a formation of a shock front at $z_l$, where the latter is 
determined by the Maxwell equal area construction: The general solution in this 
region is again of the form  (\ref{eqn:genform}) with $z_r = p$. The location $z_l$ of the shock is determined by the equal area rule as
\begin{equation}
G'[\Phi(z_l)] = \frac{G[\Phi(z_l)] - G[0]}{\Phi(z_l)}
\end{equation}
with
\begin{equation}
\Phi(z_l) = \frac{3}{2} \; \frac{q - t}{1 - 3t}
\end{equation}
so that
\begin{equation}
z_l = q - \frac{3}{4} \; \frac{(q-t)^2}{1 - 3t}.
\end{equation}

The resulting profile is given by $\Phi(z) = \Phi_{+}(z)$, for $z_l \le z \le z_r $ preceded by a shock discontinuity at $z_l$ that jumps from $F = 0$ to $F = \Phi(z_l)$.

The qualitative analysis of this solution is very similar to our previous discussion of regimes with a shock. The condition $q>t$ means that the lower rank teams will catch up with the middle rank teams. The condition $p>1/3$ gives $p>(t+p)/2$  meaning that higher point teams can decouple from this region and constitute a continuous solution to the right of the shock.

The social indices become

\bea
\omega_{C^{-}_{S}}&=&{\rm const}\times \delta(0)\;\\
\sigma_{C^{-}_{S}}&=&p-q+\frac{3}{4}\frac{(q-t)^2}{1-3t}
\eea

\noindent Again $\omega_{C^{-}}$ diverges much more strongly (and for all parameters) as opposed to the simple pole divergence we have seen for the particular case of parameters in regime $\omega_{C^{-}}$.

\noindent A representative graph including an exact numerical simulation of the microscopic model and the analytical construction is presented in Fig.~\ref{fig:cSminus}.

\begin{figure}[t]
\includegraphics[scale=0.45]{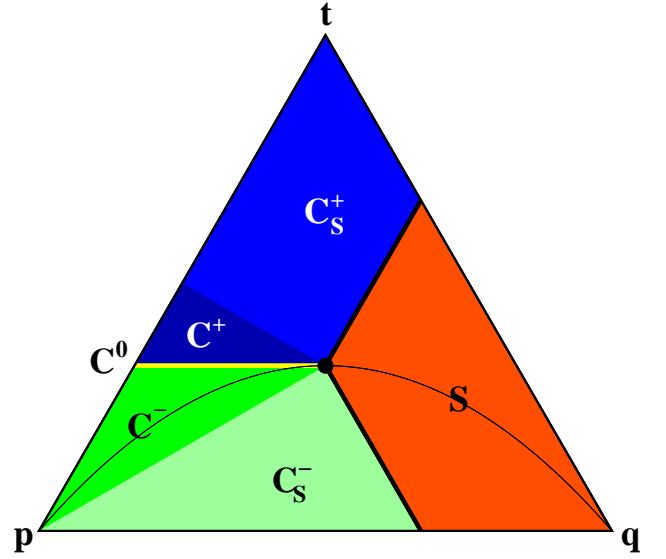}
\caption{The phase diagram of the three-agent game presented on the plane $p+t+q=1$. The blue/green areas represent the regimes with positive/negative concavity of $F$. The red area is where the solution is a single shock wave at $z=\tau/3$. The regions $C^{+}_{S}$ and $C^{-}_{S}$ are represented by the dark shades of blue/green. The crossover line from $C^{-}$ to $C^{+}$ is emphasized with a yellow line. The thin black curve represent the resolution of a three-agent game via a mini tournament between the three players in terms of two-agent competitions as described in text. }
\label{fig:phase}
\end{figure}

In Fig.~\ref{fig:phase} we have graphically combined all the regimes in a phase diagram on the $p+q+t=1$ plane . The vertices $p$, $q$ and $t$ of the triangle correspond to the cases $p=1$, $q=1$ and $t = 1$, respectively, while the edges $\overline{pt}$, $\overline{tq}$, and $\overline{qp}$, correspond to $q=0$, $p=0$ and $t=0$, respectively. The points $(p,q,t)$ with $p$ constant correspond to lines parallel to the edge $\overline{tq}$, {\it etc.}, and the dot shown represents the point $(1/3,1/3,1/3)$.

\subsection{Social Structures Emerging From Different Regimes}

If we regard the agent game as a social model where agents compete for an (unlimited) commodity through three-way competitions, we see that, depending on $(p,q,t)$, different 
wealth structures emerge asymptotically. Following an analogue of the naming scheme presented in \cite{soc5}, we can classify the resulting agent societies in terms of their wealth distributions as

\begin{itemize}
\item{} Regime $C^{-}$: Middle-class society with mild hierarchy.
\item{} Regime $C^{0}$: Pure middle-class society.
\item{} Regime $C^{+}$: Middle-class society with mild anti-hierarchy.
\item{} Regime $C^{+}_{S}$: Anti-hierarchical society.
\item{} Regime $S$: Egalitarian society.
\item{} Regime $C^{-}_{S}$: Hierarchical society.
\end{itemize}

In all regions without a shock, the agents are distributed along the same wealth span $p-q$. The reason we endow all of them with a middle-class structure is due to the fact that $F$ is a continuous function and thus has no Dirac-delta singularities in its derivative, {\it i.e.}  there are no condensations of agents at the highest or lowest points.

The regimes with a shock on the other hand constitute genuinely different societies where some agents condense at a given point accumulation rate. This implies analogies to social structures where there is a frozen class. The somewhat abused use of (anti-)hierarchy means in our context that the condensation happens in the (upper)lower part of the society. The use of the adjective egalitarian when all agents increase their points at the same rate is somewhat more natural. 

The regimes with a shock have another common feature: unlike the regimes without a shock the parameters {\bf do not} obey $p\geq t\geq q$; that is the game is not purely competitive. For instance, if the lower rank agents are favored over the higher rank ones, such as in regime $S$, we might make an analogy with a sort of welfare system which in the long run results in an equal wealth distribution. In a regime where the middle rank agents are favored over the higher rank ones, regime $C^{+}_{S}$ is suggestive of affirmative action. In regime $C^{-}_{S}$ the lower rank is favored over the middle rank and the ultimate outcome is to coalesce these regimes resulting in a large volume of poor agents. This constitutes suggestions of affirmative action where the middle class is the sole source of wealth flux to the lower class. An upper-class also persists in this regime and the resulting structure can therefore be called very hierarchical. 

However, one must be careful in that a particular $p,t,q$ ordering does not necessarily ensure that the solution will be in a unique regime, as evident from our analysis of the different regimes. For regimes with a shock, we show below whether a particular ordering resides in a single regime or not

\begin{enumerate}
\item{} $p>q>t$: Single regime: $C^{-}_{S}$.
\item{} $q>p>t$: Two regimes: $S$ or $C^{-}_{S}$.
\item{} $q>t>p$: Single regime: $S$.
\item{} $t>q>p$: Two regimes: $S$ or $C^{+}_{S}$.
\item{} $t>p>q$: Single Regime: $C^{+}_{S}$.
\end{enumerate}

\noindent An inspection of the permutation index of the orderings relative to the pure competitive ordering $p\ge t\ge q$ allows us to identify if an ordering defines a single region or not. This is given by

\be
\Pi_{o}=(-1)^{R_{o}}\;\;
\ee

\noindent with $\Pi_{o}$ defining the permutation index of the regime with a shock relative to the ones without and $R_{o}$ is the number of regimes defined by the ordering: one or two.

\subsection{Resolution in terms of a two agent game}

A three-agent game will be characterized by the set of numbers $\left(p,t,q\right)$ which controls the outcome of a single competition. However it is possible to resolve a subset of genuine three-agent games in terms of two-agent games. The simplest such approach is
to let each three teams play a single two-agent game with each other, that is to have a tournament. All the two-agent games in this tournament are decided based on how many tournaments the agents have won before. 
That is, during the tournament the tournament wins of each team is kept constant but they accumulate match points depending on the tournament wins The winner of the tournament is the agent with largest number of accumulated match points. As usual ties are decided on the basis of  equal likelihood. 
A two agent game is characterized by two numbers $p_{2}+q_{2}=1$, with $p_{2}$ denoting the probability that the team with the larger number of points, out of the two, wins. During the tournament the tournament wins of each team is, as we have pointed out, kept constant but they accumulate match points and the winner of the tournament is the agent with largest number of accumulated match points. As usual ties are decided on the basis of  equal likelihood. 
 A simple analysis yields the following probabilities to emerge as the winner out of such a tournament

\begin{subequations}
\bea
p&=&p_{2}^{2}+\frac{1}{3}p_{2}q_{2}\\
t&=&\frac{4}{3}p_{2}q_{2}\\
q&=&q_{2}^{2}+\frac{1}{3}p_{2}q_{2}
\eea 
\end{subequations}

These are represented in the phase diagram for the three-agent game in Fig.~(\ref{fig:phase}). As one might expect, for $p_{2}>1/2$ one is in the $C^{-}$ regime. The value $p_{2}=1/2$ means that $p=t=q=1/3$ and we have a completely random game represented by a shock. For values of $p_{2}$ less than $1/2$ the shock remains as in the two-agent game and in the terminology of the three-agent phase diagram we are in regime $S$. Note that the $C^{-}_S$, $C^0$, $C^+$ and 
$C^{+}_S$ regimes are absent in this version of the three agent game.
 
\subsection{Adding a decline rate for agents}

An interesting extension of the model we presented is to allow some sort of mechanism with which agents lose points. This for instance can account for the realistic observation that through inaction competitors loose fitness. A simple realization of this is to allow a {\bf steady} and democratic decline rate for agents, as 
advocated and studied in detail for two-agent games in \cite{soc5}. Such a decline rate for agents will simply result in the following generalization of (\ref{eq:cont1})

\be
\frac{\partial F}{\partial \tilde{\tau}}=-\frac{\partial F}{\partial \tilde{x}}\left[r+G'(F)\right]\;.
\ee 

\noindent where $r$ is the rate with which agents loose points. 

This new form of the equation is entirely related to the old one by a Galilean transformation

\begin{subequations}
\bea
\tilde{\tau}&=&\tau\\
\tilde{x}&=&x+r\tau
\eea
\end{subequations}

One can therefore use the solutions of the  old equation (\ref{eq:cont1}) to generate solutions for the new one by just left translating the $x$ axis. This is analogous to just left shifting the $z$ variable by $r$. However
even though the equation is Galilean invariant the boundary conditions are not: we do not allow for negative points for agents. Therefore the recipe for generating solutions for the advance-decline model is to take a solution without $r$, left translate the $z$ axis by $r$ and discard the solution for $z<0$. This will possibly generate condensation of agents at zero points, and hence a shock at $z=0$. 

Presented this way, adding a decline rate is straightforward and the regimes we have presented will in general double in number. Whether a shock at $z=0$ 
exists or not can easily be determined by comparing $r$ and $z_{l}$. If $r>z_{l}$ there will be a shock at $z=0$, if not the solution will qualitatively look the same.

As far as the decline rate is concerned, the most interesting regime is $C^{+}_{S}$ where there is originally a shock at $z_{r}$, so in this case if $r<q$ we will have two shocks. For $C^{-}_{S}$ the effect of $r$ is qualitatively immaterial since the shock is already to the left of the curve. This is also valid for regime $S$. 

In terms of the social classification we have presented, the effect of $r$ is to possibly introduce hierarchy into the society.

\section{Discussion}

We have studied the complete dynamics of the three-agent game in the 
hydrodynamical limit of large scores and large number of 
games played, by noting that the cumulative score distribution obeys a 
hyperbolic conservation law PDE, and using the method of characteristics 
to obtain analytical solutions.

The applications of this model to realistic social data could be one the most interesting ones. 
The effect of a policy reminiscent of some sort of affirmative action can also be applied in our model in that it admits three players. 

One possible extension of our model that has potentially interesting implications within the competitive subspace is the merger option of two lower ranking agents in a game. That is two agents combining forces against the more powerful opponent. In doing so they should both increase their probabilities of winning in comparison to the case without merger.

Another interesting avenue is to increase the number of attributes in choosing a winner. This will in general mean that the rate equations we have used
will involve a multi-dimensional gradient representing the different attributes. 
These extensions are currently under study. 

\acknowledgements{}

We thank K. Fortuny for the careful reading of this manuscript. The simulations needed for this work have been performed on Gilgamesh, the computer cluster of the Feza Gursey Institute.


\end{document}